\documentclass{article}

\usepackage{PRIMEarxiv}
\usepackage{pdfpages}
\usepackage[utf8]{inputenc} 
\usepackage[T1]{fontenc}    
\usepackage{url}            
\usepackage{booktabs}       
\usepackage{amsfonts}       
\usepackage{amsthm}
\usepackage{tcolorbox}
\usepackage[title]{appendix}

\usepackage{nicefrac}       
\usepackage{microtype}      
\usepackage{lipsum}
\usepackage{fancyhdr}       
\usepackage{graphicx}       
\graphicspath{{media/}}     
\usepackage{amsmath,amssymb,amsfonts}
\usepackage{graphicx,psfrag,epsf}
\usepackage{enumerate}
\usepackage{url} 
\usepackage{bm}
\usepackage{siunitx}
\usepackage{rotating}
\usepackage{xparse}
\usepackage{pdflscape} 
\usepackage{diagbox}
\usepackage[ruled,linesnumbered]{algorithm2e}
\usepackage{hyperref}
\usepackage{setspace}

\usepackage{multirow}
\def\bSig\mathbf{\Sigma}

\title{Cure Rate Joint Model for Time-to-Event Data and Longitudinal Tumor Burden with Potential Change Points
}
\author{
  Yixiang Qu$^1$, Ethan M. Alt$^1$, Weibin Zhong$^2$, Jeen Liu$^2$, Chenguang Wang$^2$, Joseph G. Ibrahim$^1$ \\
  \\
  $^1$Department of Biostatistics, University of North Carolina at Chapel Hill, Chapel Hill, NC, USA \\
  \\
  $^2$Regeneron Pharmaceuticals, Tarrytown, New York, USA
}

\begin{document}
\maketitle

\begin{abstract}
In non‐small cell lung cancer (NSCLC) clinical trials, tumor burden (TB) is a key longitudinal biomarker for assessing treatment effects. Typically, standard‐of‐care (SOC) therapies and some novel interventions initially decrease TB; however, many patients subsequently experience an increase---indicating disease progression---while others show a continuous decline. In patients with an eventual TB increase, the change point marks the onset of progression and must occur before the time of the event. To capture these distinct dynamics, we propose a novel joint model that integrates time-to-event and longitudinal TB data, classifying patients into a change-point group or a stable group. For the change-point group, our approach flexibly estimates an individualized change point by leveraging time-to-event information. We use a Monte Carlo Expectation-Maximization (MCEM) algorithm for efficient parameter estimation. Simulation studies demonstrate that our model outperforms traditional approaches by accurately capturing diverse disease progression patterns and handling censoring complexities, leading to robust marginal TB outcome estimates. When applied to a Phase 3 NSCLC trial comparing cemiplimab monotherapy to SOC, the treatment group shows prolonged TB reduction and consistently lower TB over time, highlighting the clinical utility of our approach. The implementation code is publicly available on \href{https://github.com/quyixiang/JoCuR}{https://github.com/quyixiang/JoCuR}.
\end{abstract}


\section{Introduction}
\label{sec:intro}

In oncology clinical trials, the effectiveness of new therapies like immunotherapy, targeted therapy, or chemotherapy is measured against controls or other treatments to improve patient survival and quality of life. These trials often use overall survival (OS) as the primary outcome, as it is considered the gold standard endpoint in time-to-event trials \cite{FDA_Cancer_Endpoints}. However, OS can be suboptimal because it can take a long time for patients to reach the endpoint, delaying the availability of potentially life-enhancing drugs to the market. Additionally, there are ethical concerns associated with using OS in trials. If a patient experiences disease progression, it may be unethical to continue administering an experimental treatment when a standard-of-care (SOC) option is available. To mitigate this issue, surrogate endpoints such as progression-free survival (PFS) are often used. PFS measures the time from enrollment to disease progression \emph{or} death, which is comparatively less exposed to the risk of censoring. However, PFS is essentially a measure of tumor burden (TB), as it indirectly reflects changes in tumor size or volume over time. Moreover, progression time is typically censored, as patients' tumor burden is only measured at pre-specified time periods. Instead of relying solely on PFS, directly modeling longitudinal TB data offers a more nuanced and statistically powerful approach. Longitudinal TB measurements capture the dynamic changes in tumor size, providing a richer understanding of the disease's progression and the treatment's biological impact. By leveraging longitudinal TB data, we can potentially reduce trial duration and improve the precision of treatment effect estimates, as TB trajectories often reveal patterns of response and progression that PFS alone cannot fully capture.

However, in reality, especially in the context of non-small cell lung cancer (NSCLC) treatment, modeling TB dynamics presents several challenges. First, the TB trajectory may exhibit potential change points; following an initial treatment-induced reduction, while TB often begins to increase, in some cases it instead continues to decline steadily (referred to as the \textbf{change-point group} and \textbf{stable group}, respectively, in the sequel). Second, distinguishing between these groups is critical, as a steady decline indicates no progression and, consequently, no event. Third, for subjects in the change-point group, the timing of TB increase is constrained by the upper limit of the time to progression. Fourth, the timing of change points varies among patients. Finally, patient dropouts due to progression, death, or other reasons complicate inference from longitudinal data alone, as both the group membership and boundary information for the change point may be lost in cases of censoring. Given the complexities in modeling longitudinal TB trajectories, we propose a cure rate model tailored to address these challenges. This model distinguishes between two groups: the stable group and the change-point group. For the stable group, the model functions as a longitudinal random effects model, capturing the steady decline in TB over time, with these individuals possibly considered as ``cured'' within the observation period. We refer to the ``cured'' group as the stable group in this paper to emphasize its distinct longitudinal trajectory. For the change-point group, the model integrates longitudinal and time-to-event data, forming a joint model where the time-to-event component provides additional constraint information for the random effects in the longitudinal model. Notably, our approach reverses the traditional joint modeling framework: instead of using longitudinal information to inform the survival outcome, we employ time-to-event data to set an upper bound for the TB change point, ensuring that any increase in TB occurs before the observed or inferred progression time. This novel cure rate model allows for a more flexible and accurate representation of the heterogeneous tumor dynamics observed in NSCLC patients.

The literature on joint models for longitudinal and time-to-event data is extensive, with comprehensive reviews provided in \cite{alsefri2020bayesian, kerioui2022modelling}. Most of these approaches rely predominantly on linear models for the longitudinal component \cite{wulfsohn1997joint, hsieh2006joint, crowther2013joint, brilleman2019joint}. However, many biological processes deviate significantly from linear assumptions, prompting the development of alternative approaches. For instance, \cite{desmee2017using} applied the Stochastic-Approximation Expectation Maximization (SAEM) algorithm to jointly analyze time-to-event and longitudinal data within a nonlinear mixed-effects framework in a prostate cancer setting. Several change-point models have also been proposed in the literature. For example, individual change-point models have been developed to capture significant shifts in cognitive abilities during aging \cite{vandenhoutchange2013, dominicusrandom2008}. Similarly, \cite{lange_hierarchical_1992} and \cite{kiuchi_change_1995} introduced individual-specific change-point models to describe the progression from HIV to AIDS, characterized by T4 cell counts with a change point. \cite{brilleman_bayesian_2017} applied individual-specific change points to model BMI rebound in a longitudinal study. However, these models do not address dropout and censoring, which are critical aspects of oncology clinical trials. Previous efforts to integrate time-to-event data with longitudinal change-point models include works like \cite{altzerinakou2021change, tapsoba2011joint}. However, these approaches often neglect essential biological constraints relevant to our application, such as the fact that some subjects are not at risk for a change point, and the constraint that the change point must occur before the time of progression. \cite{alt2024jointlymodelingtimetoeventlongitudinal} incorporates censoring into the model and uses time-to-event information to constrain the change point. Unlike most joint models for longitudinal ($\bm{y}$) and time-to-event ($t$) data, which model them as $f(\bm{y}) f(t|\bm{y})$, they reverse this approach by modeling $f(t) f(\bm{y}|t)$, where the upper bound for the change point is determined by censored individuals, significantly improving change-point estimation. However, their model does not account for the possibility that some individuals in the censored group may never experience a change point, which could lead to potential model mis-specification. Additionally, they could only fit their model to one arm of the data due to the possible presence of a ``cured'' population with stable longitudinal trajectories in the treatment group, limiting its applicability in comparative studies. In contrast, our approach explicitly accounts for stable individuals by incorporating a cure-rate structure, which allows for a more comprehensive analysis of treatment effects while addressing the presence of a potentially cured subgroup.

To address the complexities introduced by the biological characteristics of the disease, we introduce a two-component mixture model that effectively dichotomizes the population. For the change-point group, individuals will experience an event, and each will have an individual-specific change point that is bounded by the event time, even when the event time is latent for censored subjects. Conversely, for the stable group, no event is anticipated during the study duration, and such subjects are not at risk for a change point. Note that our model follows a \emph{cure rate} framework, although we refer to the ``cured'' group as the stable group in this paper. Our proposed model can also be viewed as a hierarchical model, where, for the stable group, it is a random effects model for the longitudinal output, and for the change-point group, it is a joint model for longitudinal and time-to-event whose longitudinal part is a change point model and time-to-event part provides the constraint information for the change point. We employ the Monte Carlo Expectation-Maximization (MCEM) method for parameter estimation. Specifically, we conduct the E-step with a Monte Carlo approach to accelerate computations by avoiding intractable numerical integrals, and in the M-step, we directly optimize those parameters with closed-form solutions and use the Limited-memory Broyden–Fletcher–Goldfarb–Shanno (L-BFGS) algorithm to optimize those without closed-form solutions. Our model is implemented in high-performance C++ code, ensuring efficient and scalable processing of complex clinical datasets.

The rest of this paper proceeds as follows. Section \ref{sec:motivate} motivates our approach with a Phase 3 oncology trial for NSCLC. Section \ref{sec:methods} introduces the proposed model, which integrates time-to-event and longitudinal TB data to account for individualized change points and stable trajectories. Section \ref{sec:algorithm} describes the MCEM algorithm for parameter estimation and further provides the formula to estimate longitudinal trajectory based on the estimated parameters. Section \ref{sec:sims} presents a simulation study evaluating the model’s performance under various scenarios and comparing it to a change-point-only model. Section \ref{sec:real_data} applies the proposed method to the EMPOWER study, demonstrating its ability to capture treatment effects on tumor progression. Finally, Section \ref{sec:discussion} concludes the paper and provides further research directions.

\section{The EMPOWER Study}
\label{sec:motivate}

To motivate our model, we introduce the EMPOWER-Lung 1 study (henceforth referred to simply as the EMPOWER study) conducted by \cite{sezer2021cemiplimab}. This multicenter, open-label, global, phase 3 clinical trial evaluates the efficacy of cemiplimab monotherapy versus standard platinum-doublet chemotherapy. The study focuses on smokers aged 18 and older with advanced or metastatic NSCLC. Cemiplimab, a programmed death-1 (PD-1) inhibitor, may be particularly effective biologically as it blocks the PD-1 receptor pathway, thereby reactivating T-cells to recognize and eliminate cancer cells, a mechanism that is highly relevant for tumors with elevated PD-L1 expression. In this study, 710 patients were randomly assigned to either the treatment group, where 356 subjects received cemiplimab, or the control group, where 354 subjects received chemotherapy. The primary endpoints were OS and PFS. 
Besides the time-to-event information, TB was also monitored, where the tumor size is defined as the sum of target lesion diameters. TB is reported as the percent change in tumor size from baseline (PCHG). In the EMPOWER study, there is domain knowledge indicating that both the treatment and control groups typically show an initial reduction in TB, then increases. However, a subset of patients may not exhibit a change point, instead showing a steady decline in TB. To capture this dynamic, we propose a modeling approach that (1) distinguishes these two groups and (2) incorporates individual-specific change points for patients in the change-point group. This framework allows us to assign distinct intercepts, change points, pre-slopes (i.e., slope before the change point), and post-slopes (i.e., slope after the change point) for patients exhibiting a change point, while modeling a consistent overall trend for patients with a steady decline.

\begin{figure}[ht]
    \centering
    \includegraphics[width=1\textwidth]{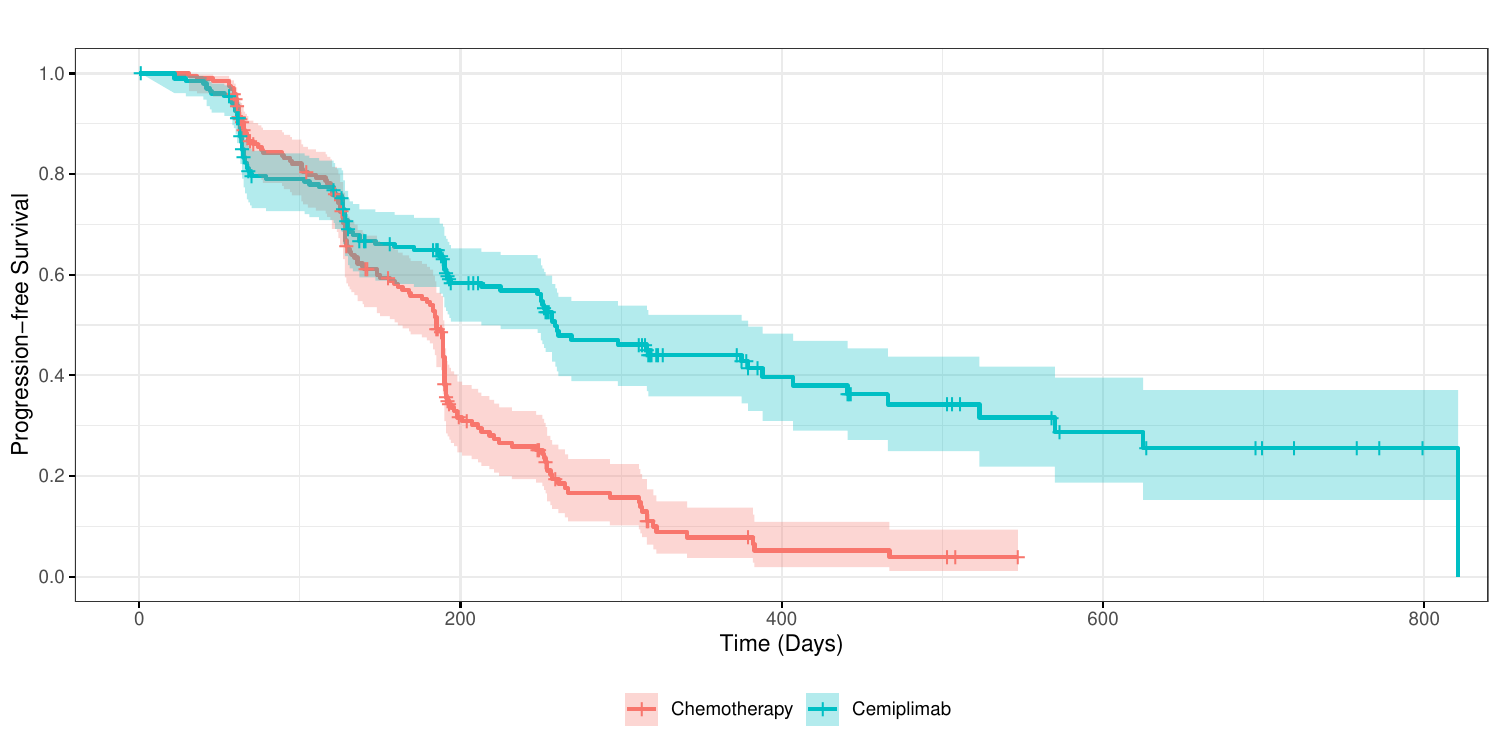}
    \caption{Kaplan-Meier curve for the two arms in the simulated EMPOWER study.}
    \label{fig:KM_2}
\end{figure}

\begin{figure}[ht]
    \centering
    \includegraphics[width=1\textwidth]{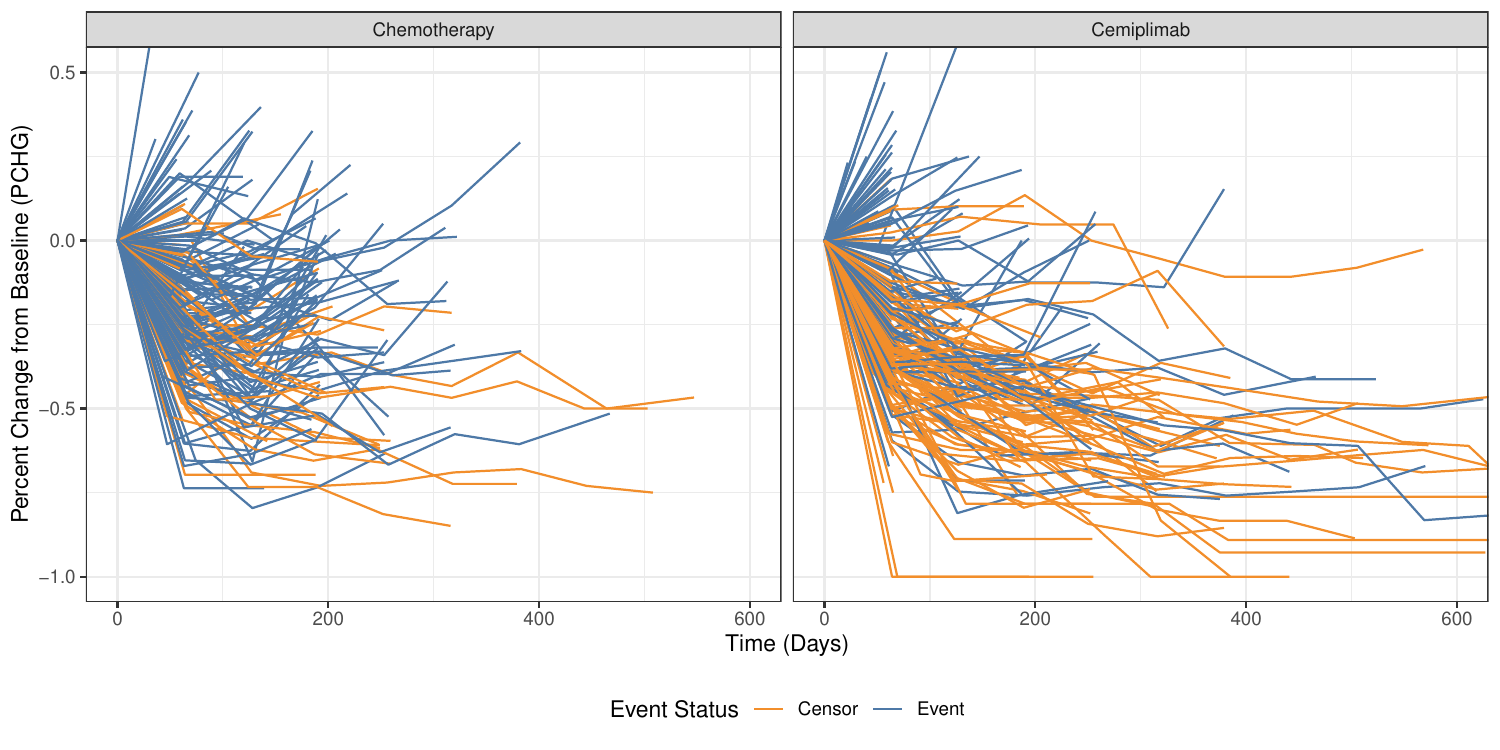}
    \caption{Longitudinal trajectories of TB among two arms in the simulated EMPOWER study.}
    \label{fig:Long_2}
\end{figure}

To further motivate our approach, we present Kaplan-Meier curves and longitudinal TB trajectories for both arms using a subset from the real data. This subset consists of a random sample of 400 out of 710 subjects (202 from the treatment arm and 198 from the control arm), selected without replacement. Although actual figures from the trial cannot be disclosed, the subset of data effectively illustrates the observed patterns and the rationale for constructing this complex model.

\begin{figure}[ht]
    \centering
    \includegraphics[width=0.6\textwidth]{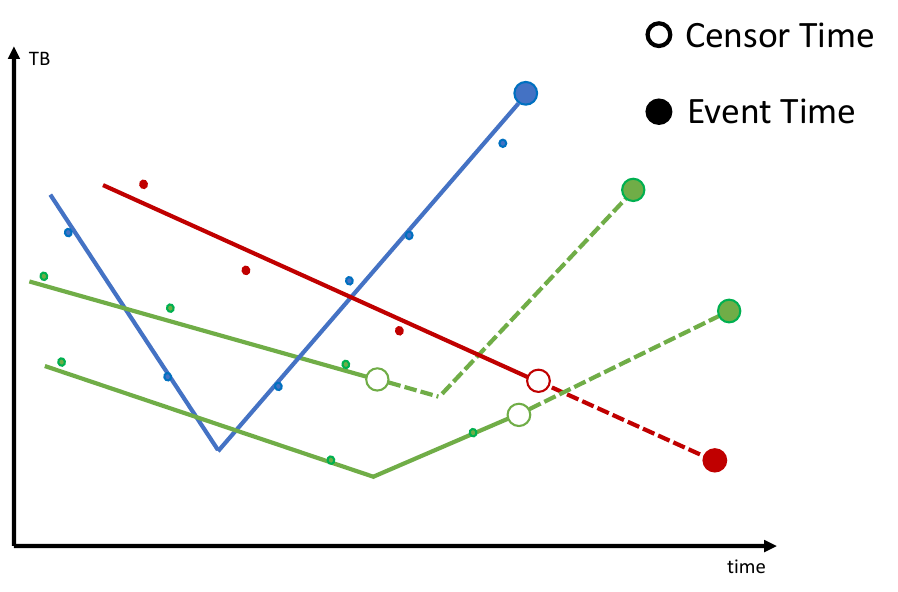}
    \caption{A simple diagram to represent the three types of data in our dataset.}
    \label{fig:three_type}
\end{figure}

Figure~\ref{fig:KM_2} presents the Kaplan-Meier curves. In the control arm (i.e., Chemotherapy), survival declines rapidly within the first 200 days and eventually approaches zero, indicating that nearly all patients experience disease progression, making the presence of a stable subgroup unlikely. In contrast, the survival curve for the treatment arm (i.e., Cemiplimab) declines more gradually and flattens after approximately 625 days, with about one quarter of patients remaining progression-free. This pattern indicates the presence of a subset of patients who are unlikely to experience a change point and may maintain a stable disease state over time---an important subgroup that must be accounted for in the model. Figure~\ref{fig:Long_2} depicts the longitudinal TB trajectories. In both arms, the subjects experiencing events show a decrease followed by an increase in TB, with timing variability indicating highly individual-specific change points. Censored subjects often show flat trajectories. But since their follow-up is truncated before an observed tumor increase, it remains unclear whether they would have eventually experienced a change point. Figure \ref{fig:three_type} provides a diagram explaining the three types of possible longitudinal trajectories in the dataset. The blue line represents subjects who experience disease progression during the follow-up period; these subjects are guaranteed to have a change point. The green and red lines represent censored subjects, distinguished by their longitudinal trajectories. Specifically, the green lines represent subjects who undergo change points similar to those shown by the blue line. Notably, the change point for these subjects can occur either before or after the censoring time, indicating the need for an individual-specific constraint on the change point derived from the time-to-event data. In contrast, the red line represents subjects in the stable group, who exhibit a decreasing longitudinal trajectory over time. Distinguishing between these two groups and identifying individual-specific change points is critical, as an overall TB trajectory cannot be established without accurately accounting for these individual variations, and our proposed model simultaneously addresses these challenges.

\section{Methods}
\label{sec:methods}

\subsection{Setup and Notation}
We consider a study with \( n \) subjects, indexed by \( i \in \{1,2,\dots,n\} \). Let $t_i^{(e)} = \min\{ t_i^{*,(e)}, c_i^{*,(e)} \}$ denote the observed progression time, where $t_i^{*,(e)}$ is the individual's event time and $c_i^{*,(e)}$ is the individual's censoring time. Let $t_i = \min\{ t_i^*, c_i^* \}$ denote the logarithm of the observed progression time, where $t_i^*$ is the log event time and $c_i^*$ is the log censoring time. Also, let $\nu_i = I( t_i \le c_i )$ denote the event indicator, where $I(A) = 1$ if $A$ is true and $0$ otherwise. Let $\bm{y}_i = (y_{i1}, \ldots, y_{i,n_i})^T$ denote the longitudinal outcome for individual $i$, where $n_i \ge 1$ is the number of longitudinal measurements recorded for subject $i$. Also, we define \( N \) as the total number of longitudinal observations across all individuals, i.e., \(N= \sum_{i=1}^n n_i \). Finally, let $\bm{s}_i = (s_{i1}, \ldots, s_{i,n_i})^T$ denote the times at which the longitudinal measurements are collected. Note that $s_{i,n_i} \le t_i^{(e)}$ since we do not observe longitudinal outcomes for individuals when they are lost to follow-up or experience disease progression or death. If someone is censored, there is a chance that this individual is in the stable group. We let $\Delta_i \in \{0, 1\}$ denote if someone is in the stable group, where an individual belongs to the stable group if $\Delta_i = 1$ and the change-point group if $\Delta_i = 0$. Note that $\Delta_i$ is \emph{partially} latent since individuals cannot belong to the stable group if they experience the event. We denote the rate of the stable individuals of the whole population by $\pi_{(s)}$, which can also be treated as ``cure rate''.

\subsection{Models}

\subsubsection{Model for stable group}

Due to the assumption that all subjects in the stable group do not experience a change point, we propose a linear mixed model for longitudinal outcomes, which can be expressed as 
%
%
$
y_{ij} = \bm{x}_{(s),ij}^T\bm{\beta}_{(s)} + b_{(s),0i} + b_{(s),1i} s_{ij} + \epsilon_{(s),ij}, \ \ j = 1, \ldots, n_i,
$, where $\bm{x}_{(s),ij}$ is a vector of covariates for subject $i$ at visit $j$ pertaining to fixed effects, $\bm{\beta}_{(s)}$ is a vector of regression coefficients with dimension $p_{(s)}$, $b_{(s),0i}$ is a random intercept, $b_{(s),1i}$ is a random slope, and $\bm{\epsilon}_{(s),i} = (\epsilon_{(s),i1}, \dots, \epsilon_{(s),i,n_i}) \sim N_{n_i}(0, \bm{\Sigma}_{\bm{y}_i | \bm{b}_{(s),i}})$ is an error term. We assume $\bm{\Sigma}_{\bm{y}_i | \bm{b}_{(s),i}} = \sigma^2_{y,(s)} \bm{I}_{n_i}$, where $\sigma^2_{y,(s)} > 0$ and $\bm{I}_n$ is the $n$-dimensional identity matrix. To simplify the notation, we define $\bm{y}_i = (y_{i1}, \ldots, y_{i,n_i})^T$, $
\bm{X}_{(s),i} =(\bm{x}_{(s),i1}, \ldots \bm{x}_{(s),i,n_i})^T
$, $\bm{b}_{(s),i} = (b_{(s),0i}, b_{(s),1i})^T$, and
$
\bm{Z}_{(s),i} =(\bm{z}_{(s),i1}, \ldots \bm{z}_{(s),i,n_i})^T
$, where $\bm{z}_{(s),ij} = (1, s_{ij})^T$ for $i = 1, \ldots, n$, $j = 1, \ldots, n_i$. The conditional distribution for the longitudinal outcome is given by
$
\bm{y}_i | \bm{b}_{(s),i} \sim N_{n_i}(\bm{X}_{(s),i} \bm{\beta}_{(s)} + \bm{Z}_{(s),i} \bm{b}_{(s),i}, \sigma^2_{y,(s)} \bm{I}_{n_i}) 
$. If a subject belongs to the stable group, then we expect the longitudinal trajectory of TB to be monotone. We thus assume that
$f_{R,(s)}( \bm{b}_{(s),i}) = \phi_2\left( \bm{b}_{(s),i} | \bm{\mu}_{r,(s)}, \bm{\Sigma}_{r,(s)} \right)$, where $\phi_n(\cdot | \bm{\mu}, \bm{\Sigma})$ is the $n$-dimensional multivariate normal distribution with mean $\bm{\mu}$ and covariance matrix $\bm{\Sigma}$, $\bm{\mu}_{r,(s)} = (\mu_{b_{(s),0}}, \mu_{b_{(s),1}})$ and $\bm{\Sigma}_{r,(s)}$ denote the mean and variance of the random effects for subjects in the stable group.

\subsubsection{Time-to-event model for the change-point group}

For the subjects in the change-point group, we propose to use a log-normal accelerated failure time (AFT) model to model the event time, i.e., we assume
%
$
\log(t_i^{*, (e)}) = t_i^{*}  = \bm{w}_i^T \bm{\gamma} + \epsilon_{tte,i}
$, where $\bm{w}_i$ represents the covariate vector for the $i$-th individual, $\bm{\gamma}$ is the $p_{tte}$ dimensional vector of coefficients corresponding to the covariates, and $\epsilon_{tte,i}$ is the error term associated with the $i$-th observation. The error term $\epsilon_{tte,i}$ is assumed to follow a normal distribution with zero mean and variance $\sigma_{tte}^2$, which captures the unexplained variability in the log-transformed event times. It is important to note that this model serves as an illustrative example, and more flexible alternative modeling approaches can also be incorporated within this framework, depending on the specific data characteristics and research objectives.

\subsubsection{Longitudinal model for change-point group}
We now discuss the longitudinal model for the change-point group. Suppose that the random effects $\bm{r}_i = (\omega_i, \bm{b}_i)$ are known, where $\omega_i$ is the subject-specific change point and $\bm{b}_i = (b_{0i}, b_{1i}, b_{2i})$ is a vector consisting of a random intercept and slopes. We propose a piecewise linear mixed model for the longitudinal outcomes with a random change point if someone is in the change-point group, i.e.,
$$
    y_{ij} = \bm{x}_{ij}^T\bm{\beta} + b_{0i} + b_{1i}(s_{ij} - \omega_{i})I( s_{ij} \le \omega_i ) + b_{2i}(s_{ij} - \omega_i)I (s_{ij} > \omega_i ) + \epsilon_{ij}, \ \ j = 1, \ldots, n_i,
$$
where $\bm{x}_{ij}$ is a $p_{long}$ dimensional vector of covariates for subject $i$ at visit $j$ pertaining to fixed effects, $\bm{\beta}$ is a vector of regression coefficients, $b_{0i}$ is a random intercept, $b_{1i}$ is a random slope prior to the change point $\omega_i$ (i.e., the pre-slope), $b_{2i}$ is a random slope after the change point (i.e., the post-slope), and $\bm{\epsilon}_i = (\epsilon_{i1}, \dots, \epsilon_{i,n_i})  \sim N_{n_i}(0, \bm{\Sigma}_{\bm{y}_i | \bm{r}_i})$ is an error term. We assume $
\bm{\Sigma}_{\bm{y}_i | \bm{r}_i} = \sigma^2_y \bm{I}_{n_i}$, where $\sigma^2_y > 0$. To simplify the notation, we define $\bm{X}_i = (\bm{x}_{i1}, \ldots, \bm{x}_{i,n_i})^T$. And we let $\Delta_{ij} = s_{ij} - \omega_i$ and let $\bm{z}_{ij} = (1, \Delta_{ij} \cdot I ( \Delta_{ij} \le 0 ), \Delta_{ij} \cdot I( \Delta_{ij} > 0 ))^T$. Moreover, let $\bm{Z}_i = (\bm{z}_{i1}, \ldots, \bm{z}_{i,n_i})^T$ denote the ``design matrix'' for the random effects $\bm{b}_i$, which itself depends on a random effect $\omega_i$. 
Therefore, if the event time is observed, the conditional distribution for the longitudinal outcome is given by
$
    \bm{y}_i | \bm{r}_i \sim N_{n_i}(\bm{X}_i \bm{\beta} + \bm{Z}_i \bm{b}_i, \sigma^2_y \bm{I}_{n_i})
$, where $\bm{Z}_i$ depends on $\omega_i$, and $\omega_i \in (0, e^{t_i^*})$. For censored individuals in the change-point group, even though $t_i^*$ is unobserved, its distribution is specified through the time-to-event model. By constraining $\omega_i$ to lie in $(0, e^{t_i^*})$, the model leverages the time-to-event information to inform the random effects, thereby enhancing the longitudinal model. In this way, our approach reverses the traditional joint modeling framework by using time-to-event data to inform the longitudinal component rather than the usual reverse.

In practice, the random effects are unobserved, and we must make distributional assumptions. We assume that the density of the random effects is given by a partially truncated multivariate normal (PTMVN) distribution, i.e., 
$
    f_{R|T}( \bm{r}_i | t_i^{*}) \propto \phi_4\left( \bm{r}_i | \bm{\mu}_r, \bm{\Sigma}_r \right) I( 0 < \omega_i \le e^{t_i^*} ),
      %
$
where $\bm{r}_i = ( \omega_i, \bm{b}_i^T)^T$ denotes the random effects for subject $i$ who does not experience a change point. Additionally, $\bm{\mu}_r = (\mu_{\omega}, \bm{\mu}_b^T)^T$ and 
$\bm{\Sigma}_r = 
    \begin{pmatrix} 
        \sigma_{\omega}^2   & \bm{\sigma}_{\bm{b}\omega}^T \\
        \bm{\sigma}_{\bm{b}\omega} & \bm{\Sigma}_{\bm{b}} 
    \end{pmatrix}
$ 
are, respectively, the mean and covariance matrix of the random effects prior to truncation. Properties of the PTMVN distribution are studied in \cite{alt2024jointlymodelingtimetoeventlongitudinal}. Henceforth, to simplify the notation, we use $PTMVN (\bm{\mu}_r, \bm{\Sigma}_r, 0, e^{t_i^*})$ to denote the distribution of $f_{R|T}$ in this paper.

\subsection{Complete data likelihood}

We first introduce parameters for the distribution corresponding to the change-point group (i.e., those with $\Delta_i = 0$). Let the parameters for the models of the time-to-event outcome, random effects, and longitudinal outcome be denoted respectively by $\bm{\theta}_T$, $\bm{\theta}_R$, and $\bm{\theta}_Y$, where in our applications $\bm{\theta}_T$ includes the regression coefficients $\bm{\gamma}$ and the variance $\sigma_{tte}^2$ from the log‐normal AFT model; $\bm{\theta}_R$ includes the mean vector $\bm{\mu}_r = (\mu_{\omega}, \bm{\mu}_b^T)^T$ and the covariance matrix $\bm{\Sigma}_r$; and $\bm{\theta}_Y$ includes the fixed‐effects coefficients $\bm{\beta}$ and the error variance $\sigma_y^2$ in the longitudinal change‐point model. Moreover, let $f_T(\cdot| \bm{\theta}_T)$, $f_{R|T}(\cdot| \bm{\theta}_R)$, and $f_{Y|R}(\cdot| \bm{\theta}_Y)$ denote the parametric densities proposed for the time-to-event outcome, the random effects (conditional on the event time), and the longitudinal outcomes (conditional on the random effects), respectively. On the other hand, for patients not at risk for the change point, we denote the parameters for the random effects and longitudinal outcome by $\bm{\theta}_{R,(s)}$ and $\bm{\theta}_{Y,(s)}$, respectively. In our application, $\bm{\theta}_{R,(s)}$ includes the mean vector $\bm{\mu}_{r,(s)}$ and the covariance matrix $\bm{\Sigma}_{r,(s)}$ of the random effects for the stable group, and $\bm{\theta}_{Y,(s)}$ includes the fixed‐effects coefficients $\bm{\beta}_{(s)}$ and the error variance $\sigma_{y,(s)}^2$ in the longitudinal model for stable subjects. Similarly, we use $f_{R,(s)}(\cdot| \bm{\theta}_{R,(s)})$ and $f_{Y|R,(s)}(\cdot| \bm{\theta}_{Y,(s)})$ to denote the corresponding densities. Let $\bm{\theta} = \left(\pi_{(s)}, \bm{\theta}_T^T,\, \bm{\theta}_R^T,\, \bm{\theta}_Y^T,\, \bm{\theta}_{R,(s)}^T,\, \bm{\theta}_{Y,(s)}^T\right)^T$ denote all model parameters.
%
The complete data likelihood is given by
\begin{align}
    L^c(\bm{\theta} | D^c) & \propto  
        \prod_{i=1}^{n} \left[ (1- \pi_{(s)})  f_T( t_i| \bm{\theta}_T) \cdot  f_{R|T}( \bm{r}_i | t_i, \bm{\theta}_R)\cdot  f_{Y|R}(\bm{y}_i | \bm{r}_i, \bm{\theta}_Y) \cdot I(t_i^* = t_i)
        \right]^{\nu_i}
    \notag \\
    & \times 
        \prod_{i=1}^{n} \left[ 
        (1- \pi_{(s)})
        f_T(t_i^* | \bm{\theta}_T) \cdot f_{R|T}( \bm{r}_i | t_i^*, \bm{\theta}_R)\cdot  f_{Y|R}(\bm{y}_i | \bm{r}_i, \bm{\theta}_Y) \cdot I(t_i^* > t_i )\right]^{(1-\nu_i)(1-\Delta_i)} 
            \notag \\
    & \times 
        \prod_{i=1}^{n} \left[
        \pi_{(s)}
        f_{R,(s)}( \bm{r}_{(s),i} |  \bm{\theta}_{R,(s)})\cdot  f_{Y|R,(s)}(\bm{y}_i | \bm{r}_{(s),i}, \bm{\theta}_{Y,(s)}) \cdot I(t_i^* > t_i) \right]^{(1-\nu_i)\Delta_i },
    \label{eq:like_complete}
\end{align}
where $\bm{r}_{(s),i} = \bm{b}_{(s),i}$ denotes the random effects for subject $i$ who is in the stable group. Each row in \eqref{eq:like_complete} represents a subpopulation in the dataset. In particular, the first row corresponds to the complete data likelihood for the individuals who experience the event, the second row corresponds to the complete data likelihood for individuals who are censored and experience the change point, and the final row corresponds to the complete data likelihood for subjects who are censored and belong to the stable group.

\section{Algorithm}
\label{sec:algorithm}

The missing data situation in \eqref{eq:like_complete} is highly complex. For each individual, the random effects (\(\bm{r}_i\) or \(\bm{r}_{(s),i}\)) are unobserved. In addition, for censored individuals, both the event time \(t_i^*\) and the stable group indicator \(\Delta_i\) are also latent. Due to this intricate structure of missing variables, the EM algorithm is a natural choice for parameter estimation since it provides a systematic and iterative approach to handling these unobserved components. To infer the maximum likelihood estimations (MLEs) of the parameters in \eqref{eq:like_complete}, we first derive the Q function. Taking the expectation of the logarithm of \eqref{eq:like_complete} and rearranging the terms, we obtain the Q function, which is shown explicitly in Section 1.1 of the Supplementary Note. Due to its length, we decompose it into four components, i.e., $Q = Q^{(1)} + Q^{(2)} + Q^{(3)} + Q^{(4)}$. Henceforth, to simplify our notation, we use subscript 0 or 1 to denote $\Delta=0$ or $\Delta=1$, respectively, and we use superscript $(=)$ or $(>)$ to denote the situation where $t_i^* = t_i$ or $t_i^* > t_i$. All expectations in this section are conditional on the observed data. In particular, we let $E^{(=)}[A_i]$ denote $E[A_i |t_i^* = t_i, \bm{y}_i]$, $E^{(>)}[A_i]$ denote $E[A_i |t_i^* > t_i, \bm{y}_i]$, $E_0^{(=)}[A_i]$ denote $E[A_i |t_i^* = t_i, \bm{y}_i, \Delta_i = 0]$, $E_0^{(>)}[A_i]$ denote $E[A_i |t_i^* > t_i, \bm{y}_i, \Delta_i = 0]$, and $E_1^{(>)}[A_i]$ denote $E[A_i |t_i^* > t_i, \bm{y}_i, \Delta_i = 1]$.

\subsection{E-step}

\subsubsection{The terms related to $\pi_{(s)}$}


We first focus on calculating the terms related to the parameter $\pi_{(s)}$, which is represented by $Q^{(1)}$ in Section 1.2 of the Supplementary Note. Since $\Delta_i$ is a binary random variable, a straightforward calculation using Bayes' theorem gives
$
    {E}^{(>)}[\Delta_i]
    = {P}(\Delta_i =1 |t_i^* > t_i, \bm{y}_i)=\frac{{P}(\bm{y}_i| \Delta_i=1) \pi_{(s)}}{{P}(\bm{y}_i| \Delta_i=1) \pi_{(s)}+ {P}(t_i^*>t_i,\bm{y}_i| \Delta_i=0) (1-\pi_{(s)})}
$. The calculation of it requires the calculation of ${P}(\bm{y}_i| \Delta_i=1)$ and ${P}(t_i^*>t_i,\bm{y}_i| \Delta_i=0)$, and the details are provided in Section 2.2 of the Supplementary Note.

\subsubsection{Subjects whose event times are observed}

The terms representing those subjects whose event times are observed are given as $Q^{(2)}$ in Section 1.2 of the Supplementary Note. The expectation in $Q^{(2)}$ is approximated using Monte Carlo integration. All the quantities to be calculated in $Q^{(2)}$ can be expressed as ${E}_0^{(=)}[h_1(t_i^*) \cdot h_2(\omega_i) \cdot h_3(\bm{b}_i)]$, where $h_1$, $h_2$ and $h_3$ denote any arbitrary functions. To calculate it, we may rewrite it as \eqref{eq:h123}. The Monte Carlo procedure to approximate \eqref{eq:h123} and the proof of it are provided in Section 2.3 and Section 7 of the Supplementary Note, separately.
\begin{align}
&{E}^{(=)}_0[h_1(t_i^*) \cdot h_2(\omega_i) \cdot h_3(\bm{b}_i)] \notag\\ &= \frac{\int_0^{e^{t_i^*}} h_1(t_i^*) \cdot h_2(\omega_i) {E} [h_3(\bm{b}_i)| \bm{y}_i, \omega_i , \Delta_i = 0]
 f(\bm{y}_i |\omega_i, \Delta_i = 0) f( \omega_i| t_i^*, \Delta_i = 0) d\omega_i}{\int_0^{e^{t_i^*}} f(\bm{y}_i |\omega_i, \Delta_i = 0) f( \omega_i| t_i^*, \Delta_i = 0) d\omega_i}.
 \label{eq:h123}
\end{align}

\subsubsection{Censored subjects who experience change point}

The terms related to those who are censored and experience a change point are written as $Q^{(3)}$ in Section 1.2 of the Supplementary Note.
Similar to \eqref{eq:h123}, for the quantity in $Q^{(3)}$, the E-step for $h_1(t_i^*) \cdot h_2(\omega_i) \cdot h_3(\bm{b}_i)$ is shown in \eqref{eq:h123_cen}. The Monte Carlo procedure to approximate \eqref{eq:h123_cen} and the proof of it are provided in Section 2.4 and Section 7 of the Supplementary Note, separately.
\begin{align}
&{E}^{(>)}_0[h_1(t_i^*) \cdot h_2(\omega_i) \cdot h_3(\bm{b}_i)]\notag\\& = \frac{\int_{t_i}^\infty \int_0^{e^{t_i^*}}h_1(t_i^*) \cdot h_2(\omega_i) {E}[ h_3(\bm{b}_i) | \omega_i, \bm{y}_i, \Delta_i=0] f(\bm{y}_i|\omega_i, \Delta_i=0) f( \omega_i| t_i^* , \Delta_i=0) f(t_i^* | \Delta_i=0)  d\omega_i d{t_i^*}}{\int_{t_i}^\infty \int_0^{e^{t_i^*}}  f(\bm{y}_i|\omega_i, \Delta_i=0) f( \omega_i| t_i^*, \Delta_i=0) f(t_i^*|\Delta_i=0)  d\omega_i d{t_i^*}}.
\label{eq:h123_cen}
\end{align}

\subsubsection{Censored subjects who are in the stable group}

Lastly, we use $Q^{(4)}$ defined in Section 1.2 of the Supplementary Note to represent those censored subjects who are not at risk of change points. In $Q^{(4)}$, we note that there are two quantities to be calculated in the E-step, namely ${E}_1[\bm{b}_{(s),i}]$ and ${E}_1[(\bm{b}_{(s),i}-\bm{\mu}_{r,(s)})^{\otimes 2}]$. And they can both be expressed in closed form based on the properties of the multivariate normal distribution. The details of these equalities are provided in Section 2.5 of the Supplementary Note.


\subsection{M-step}

Most of the parameters in the M-step have closed-form solutions, including the cure rate, \(\pi_{(s)}\); the four parameters for the change-point group, namely \(\bm{\gamma}\), \(\sigma_{tte}\), \(\bm{\beta}\), and \(\sigma_{y}\); and the parameters for the stable group, namely \(\bm{\mu}_{r,(s)}\), \(\bm{\Sigma}_{r,(s)}\), \(\bm{\beta}_{(s)}\), and \(\sigma_{y,(s)}\). The closed-form solutions for these parameters are presented in Section 3.1 in the Supplementary Note, and the scalar forms are further provided in Section 3.2 of the Supplementary Note for easier implementation.

Two parameters, $\bm{\mu}_r$ and $\bm{\Sigma}_r$, in the M-step cannot be expressed in closed form, so we update them using the L-BFGS algorithm with box constraints. This method requires calculating the gradient with respect to both $\bm{\mu}_r$ and $\bm{\Sigma}_r$. The gradient for the vector $\bm{\mu}_r$ is relatively straightforward to compute, which is shown in Section 3.3.1 of the Supplementary Note. However, $\bm{\Sigma}_r$ is a $4 \times 4$ positive-definite matrix with 10 degrees of freedom (fewer than the 16 elements in the matrix). To optimize $\bm{\Sigma}_r$, we propose to use Cholesky decomposition, expressing $\bm{\Sigma}_r$ as $\bm{\Sigma}_r = \bm{O}_r \bm{O}_r^T$, where $\bm{O}_r$ is a lower triangular matrix. Since the diagonal entries of $\bm{O}_r$ must be positive, we reparameterize $\bm{O}_r$ as $\bm{P}_r$, where $(\bm{P}_r)_{ij} = (\bm{O}_{r})_{ij}$ if $i \ne j$ and $(\bm{P}_r)_{ii} = \log\left\{ (\bm{O}_{r})_{ii} \right\}$ otherwise. We then compute the gradient of the objective function \(Q\) with respect to $\text{vech}(\bm{P}_r)$, i.e., $
\frac{\partial Q}{\partial \operatorname{vech}(\bm{P}_r)}
$. To derive this gradient, we first apply the chain rule to express it as $\frac{\partial Q}{\partial \text{vech}(\bm{P}_r)} = \frac{\partial \text{vech}(\bm{O}_r)^T}{\partial \text{vech}(\bm{P}_r)} \frac{\partial Q}{\partial \text{vech}(\bm{O}_r)} = \frac{\partial \text{vech}(\bm{O}_r)^T}{\partial \text{vech}(\bm{P}_r)}
    \frac{\partial \text{vech}(\bm{\Sigma}_r)^T}{\partial \text{vech}(\bm{O}_r)} \frac{\partial Q}{\partial \text{vech}(\bm{\Sigma}_r)}$, and then we calculate $\frac{\partial \text{vech}(\bm{O}_r)^T}{\partial \text{vech}(\bm{P}_r)}
$, $\frac{\partial \text{vech}(\bm{\Sigma}_r)^T}{\partial \text{vech}(\bm{O}_r)}$, and $\frac{\partial Q}{\partial \text{vech}(\bm{\Sigma}_r)}$ separately, whose details are provided in Section 3.3.2 of the Supplementary Note.
%

\subsection{Uncertainty quantification}






We propose to use the bootstrap method to account for uncertainty (e.g., the computation of confidence intervals (CIs)) in the estimation for the model parameters. Specifically, the bootstrap procedure involves resampling the original dataset with replacement, drawing the same number of subjects as in the original dataset, and repeating this process $B$ times, where we let $B=500$ in our simulation and real data application. For each bootstrap sample, we re-estimate the model parameters using the proposed MCEM algorithm, resulting in $B$ estimates for each parameter of interest. A 95\% confidence interval for each parameter is computed using the empirical distribution of $B$ bootstrap estimates, with the lower and upper bounds defined by the 2.5th and 97.5th percentiles, respectively.

\subsection{Longitudinal trajectory estimation}
In line with the estimands framework \cite{kahan2024estimands}, which emphasizes modeling marginal effects, quantifying the marginal expectation of TB at a given time point is crucial for deriving causal contrasts such as the average treatment effect. In this section, we describe our method for estimating the \emph{marginal} longitudinal trajectory of TB.


It can be shown that the marginal TB mean at a given time point \( s \) is given by
%
%
\begin{align}
    E[y |  s , \bm{{\theta}}]
    = {\pi}_{(s)} {\bm{\mu}}_{(s)}( s ) + (1 - {\pi}_{(s)}) {\bm{\mu}}_{(cp)}( s ),
    \label{eq:E_y_final}
\end{align}
where ${\bm{\mu}}_{(s)}(s) = E[\bm{X}_{(s)}] {\bm{\beta}}_{(s)} + {\mu}_{b_{(s),0}} + {\mu}_{b_{(s),1}} s$ is the marginal mean at the new time $s$ for the stable group and ${\bm{\mu}}_{(cp)}(s) = E[\bm{X}] {\bm{\beta}} + E[{b}_{0} + b_{1}(s - \omega)I( s\le \omega ) + b_{2}(s - \omega)I( s > \omega )|  \bm{{\theta}}]$ is the marginal mean at the new time $s$ for the change point group. The derivation of \eqref{eq:E_y_final} is presented in Section 4.1 in the Supplementary Note. To obtain the marginal TB based on the estimated parameters $\hat{\bm{\theta}}$, we can substitute them into the equation. We further note that the calculation of ${\bm{\mu}}_{(cp)}(s)$ requires the calculation of $E[{b}_{0} + b_{1}(s - \omega)I( s\le \omega ) + b_{2}(s - \omega)I( s > \omega )|  \bm{{\theta}}]$, which we propose to obtain via Monte Carlo sampling, whose details are presented in Section 4.2 in the Supplementary Note. Additionally, in our application, the design matrix consists only of measurements at baseline. Thus, we may substitute $E[\bm{X}]$ with $\bar{\bm{X}} := n^{-1} \sum_{i=1}^n \bm{X}_i$, which is a consistent estimator for $E[\bm{X}]$ by the weak law of large numbers. 
More generally, some components of $\bm{X}_i$ could depend on time, in which case some measurements would be missing for subjects who died or were censored. In these cases, one may generalize the development by considering multiple imputation approaches or modeling the joint distribution of the covariates and the time-to-event and longitudinal outcomes.

\begin{algorithm}[ht]
\caption{Estimation Procedures of the Marginal Longitudinal Trajectory}
\label{alg:trajectory}
\KwIn{A new time point $s$ and estimated parameters ${\bm{\theta}}$.}
\KwOut{Estimated marginal longitudinal outcome ${E}[y|s, \bm{{\theta}}]$.}
\BlankLine
Initialize $S_{\text{cp}} \gets 0$\;
\For{$j \in \{1,2, \cdots, J\}$}{
    Sample $t^{(j)} \sim {N}\bigl(\bar{\bm{w}}^T {\bm{\gamma}},\, {\sigma}^2_{tte}\bigr)$\;
    Sample $(\omega^{(j)}, b_0^{(j)}, b_1^{(j)}, b_2^{(j)})^T \sim PTMVN\Bigl({\bm{\mu}}_r,\, {\bm{\Sigma}}_r,\, 0,\, e^{t^{(j)}}\Bigr)$\;
    
    $
    y_{\text{cp}}^{(j)} \gets b_0^{(j)} + b_1^{(j)}\,(s - \omega^{(j)})\,I\bigl(s \le \omega^{(j)}\bigr) + b_2^{(j)}\,(s - \omega^{(j)})\,I\bigl(s > \omega^{(j)}\bigr)
    $\;
    $S_{\text{cp}} \gets S_{\text{cp}} + y_{\text{cp}}^{(j)}$\;
}

${\bm{\mu}}_{(cp)}(s)  \gets \bar{\bm{X}}\,{\bm{\beta}} + \frac{S_{\text{cp}}}{J}$\;

$
{\bm{\mu}}_{(s)}(s) \gets \bar{\bm{X}}_{(s)}\,{\bm{\beta}}_{(s)} + {\mu}_{b_{(s),0}} + {\mu}_{b_{(s),1}}\,s$\;

${E}[y|s,\bm{{\theta}}] \gets {\pi}_{(s)}\,{\bm{\mu}}_{(s)}(s) + \bigl(1 - {\pi}_{(s)}\bigr)\,{\bm{\mu}}_{(cp)}(s) $\;

\Return{${E}[y|s,\bm{{\theta}}]$.}
\end{algorithm}

We summarize the entire procedure of calculating \eqref{eq:E_y_final} in Algorithm \ref{alg:trajectory}. And in order to construct a 95\% CI for the marginal longitudinal trajectory at the new time point \( s \), we employ a bootstrap procedure as follows. For each bootstrap iteration \( b \in \{1, 2, \ldots, B\} \), a resample is drawn from the original dataset with replacement and the MLEs \({\bm{\theta}}_b\) are computed using the proposed MCEM algorithm. Then, using Algorithm~\ref{alg:trajectory}, we obtain the bootstrap estimate \(E_b[y | s, {\bm{\theta}}_b]\) of the marginal longitudinal trajectory. After obtaining \(B\) such estimates, we sort them and take the 2.5th and 97.5th percentiles as the lower and upper bounds, respectively, of the 95\% CI for the marginal trajectory.

To quantify the difference in TB between the treatment and control arms, we estimate the \emph{average treatment effect} (ATE) at any time $s$ as \(
\text{ATE} = E[y | s, \bm{\theta}_{A = 1}] - E[y | s, \bm{\theta}_{A = 0}]
\), where \( \bm{\theta}_{A = 1} \) represents the estimated parameters for the treatment arm, and \( \bm{\theta}_{A = 0} \) represents the estimated parameters for the control arm. Given our approach to estimate the marginal mean for each arm, the ATE can be directly obtained by taking their difference. A 95\% CI for the ATE can be constructed using the bootstrap method by resampling and computing the percentile-based CI from the bootstrap estimates.

\section{Simulation study}
\label{sec:sims}

\subsection{Simulation procedure}

In this simulation study, we aim to replicate key aspects of the real EMPOWER study. The parameters used for generating the data are meticulously chosen to mimic the trial's characteristics. Those parameters are listed in Section 5.1 of the Supplementary Note.


Consider a simulation setup with $n$ subjects. For $i \in \{1,2, \ldots, n \}$, we first simulate the progression times $t_i^{*,(e)}$ using a log-normal AFT model. Specifically, the log-normal density function for the progression times is given by $
f(t_i^{*,(e)}) = \frac{1}{t_i^{*,(e)} \sigma_{tte} \sqrt{2\pi}} \exp \left(-\frac{(\log t_i^{*,(e)} - {\omega}_i {\gamma}_1)^2}{2\sigma_{tte}^2}\right)
$, where ${\gamma}_1$ is the regression coefficient and $\sigma_{tte}$ is the standard deviation of the logarithmic transformation of the progression times. We generate ${\omega}_i$ with a standard normal distribution to reflect the fact that the covariate is a scaled baseline in our application. Following the generation of progression times, each subject is randomly assigned to either a stable or a change-point group based on a probability $\pi_{(s)}$. Subjects in the stable group are reassigned with an infinite progression time, $t_i^{*,(e)} = \infty$, representing no change point. Censoring times $c_i^{*,(e)}$ are independently generated from an exponential distribution with a fixed rate parameter. 
The observed event times are then determined by $t_i^{(e)} = \min\{t_i^{*,(e)}, c_i^{*,(e)}\}$.

For those subjects who are assigned to the change-point group, we generate their random effects and longitudinal outcomes as follows. We first generate random effects $ (\omega_i, \bm{b}_i^T)^T$ using the PTMVN distribution such that $f(\omega_i, \bm{b}_i) \propto N(\bm{\mu}_r, \bm{\Sigma}_r) I(\omega_i < t_i^{*,(e)})$. Then, for each subject $i=1,\ldots,n$ and each visit $j=1,2,\ldots,J_n$, where $J_n$ is a sufficiently large constant, visit times are generated according to
$
s_{ij}^* = \left|0.1 \times j - z_{ij}\right|
$, with $z_{ij} \sim \text{Half-Normal}(0, 0.02^2)$ introducing variability such that $E[s_{ij}^*] \approx 0.10 \times (j-1) + 0.08$ years and $\operatorname{Var}(s_{ij}^*) \approx 0.012^2$ years. The observed visit times for subject $i$, $s_{ij}$, are taken to be all visit times prior to the observed event time, i.e., we calculate $n_i = \max\{ j : s_{ij}^* \le t_i^{(e)} \}$ and take $s_{ij} = s_{ij}^*$ for $j = 1, \ldots, n_{i}$. If subjects progressed or dropped out prior to the first observed visit time, i.e., if $t_i^{(e)} < s_{i1}$, we set $s_{i1} = 0.1 \times s_{i1}^*$. The longitudinal outcome is then generated via
$
    y_{ij} \sim N\left( \bm{x}_i{\beta}_1 + b_{0i} + b_{1i}(s_{ij} - \omega_i) + b_{2i}(s_{ij} - \omega_i), \sigma^2_y \right)
$
for $i = 1, \ldots, n$ and $j = 1, \ldots, n_i$. Furthermore, $\bm{x}_i$ is constructed by stacking $x_i$ with $n_i$ times, where $x_i$ follows an arbitrary distribution. Specifically in this simulation, we generate $x_i$ using a standard normal distribution to mimic our real application. For those subjects who are assigned to the stable group, we first generate random effects using $(b_{(s),0i}, b_{(s),1i})^T \sim N_2(\bm{\mu}_{r,(s)}, \bm{\Sigma}_{r,(s)})$. Then we generate visit times ($s_{ij}$) using the same method as that in the change-point group. Finally, we generate the longitudinal outcome via
$
    y_{ij} \sim N\left( \bm{x}_{i}{\beta}_{(s),1} + b_{(s),0i} + b_{(s),1i} s_{ij} , \sigma^2_{y,(s)} \right)
$
for $i = 1, \ldots, n$ and $j = 1, \ldots, n_i$.

For $n  \in \{ 100, 200, 500 \}$ and $\pi_{(s)}\in \{0, 0.2, 0.4\}$, we evaluate the model under different settings for the longitudinal trajectory. Specifically, we assess its sensitivity to variations in pre-slope and post-slope differences using $\bm{\mu}_{r} = (0.5, 0.0, -0.5, 0.5)^T$ and $\bm{\mu}_{r} = (0.3, 0.0, -0.3, 0.3)^T$. For each scenario, we repeat the generative process $B$ times to assess the model’s performance. We compute the bias, mean squared error (MSE), and 95\% CI coverage (Cover). For a parameter $\theta$, we compute these quantities as 
$\text{Bias} = B^{-1} |\sum_{b=1}^B (\hat{\theta}_b - \theta )|$, 
$\text{MSE} = B^{-1} \sum_{b=1}^B ( \hat{\theta}_b - \theta )^2$, and 
$\text{Cover} = B^{-1} \sum_{b=1}^B I( L(\theta | D_b) \le \theta \le U(\theta | D_b) )$, where $D_b$ refers to the $b^{th}$ generated data set, $\hat{\theta}_b$ is the estimate of $\theta$ from $D_b$, $L(\theta | D_b)$ and $U(\theta | D_b)$ are, respectively, the 2.5\% and 97.5\% empirical quantiles of $\hat{\theta}_b$ for data set $D_b$. In addition to evaluating the parameter estimates, we also assess the estimation of the longitudinal trajectory. The ground truths for longitudinal trajectory under each scenario are obtained by averaging over 2000 simulations, with time points ranging from 0.1 to 2 in increments of 0.1. The estimated longitudinal trajectory is then computed using Algorithm \ref{alg:trajectory}. And its bias, MSE, and Cover are calculated in the same manner as described above.

We compare our model with another model that assumes that everyone experiences a change point proposed by \cite{alt2024jointlymodelingtimetoeventlongitudinal}, which is a special case of our model when $\pi_{(s)}=0$. We repeat the generative process $500$ times and evaluate the parameter estimations as well as the marginal longitudinal trajectory estimations.

\subsection{Simulation results}

\begin{sidewaysfigure}
    \centering
    \includegraphics[width=\linewidth]{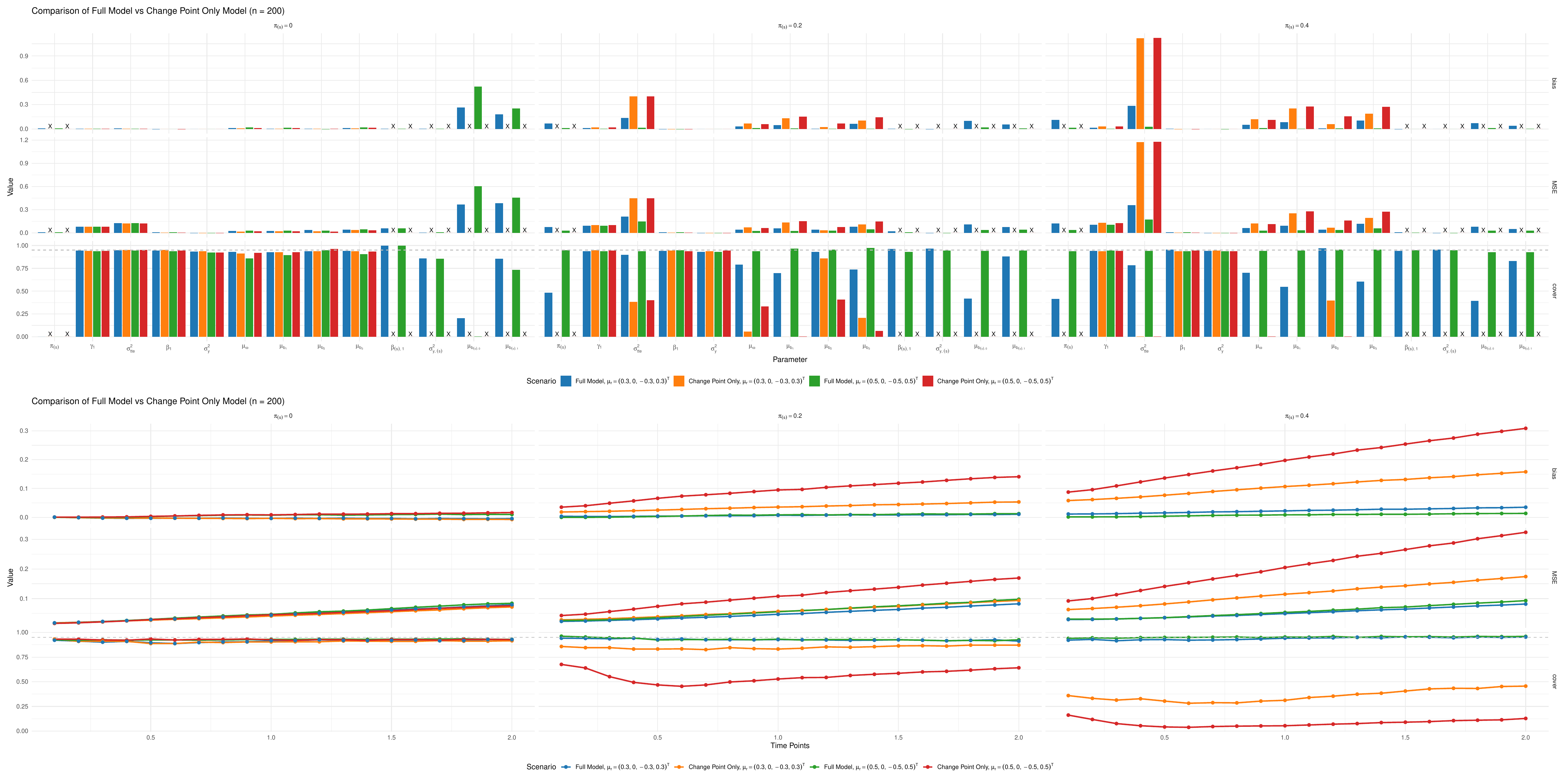}
    \caption{Evaluations for Parameter Estimations and Longitudinal Trajectory Estimations Under Different Scenarios ($n=200$). The ``X'' marks indicate the absence of specific parameters in certain models.}
    \label{fig:sim_res}
\end{sidewaysfigure}

We first provide the simulation results when $n=200$ in Figure \ref{fig:sim_res}, while the results for $n=100$ and $n=500$, which exhibit similar patterns, are provided in Section 5.2 of the Supplementary Note. The simulation results for parameter estimation when $n=200$ are presented in the top part of Figure \ref{fig:sim_res}. We first assess the model's performance under conditions where the pre-slope and post-slope parameters exhibit a relatively large divergence, specifically setting $\bm{\mu}_{r} = ( 0.5, 0.0, -0.5, 0.5)^T$. These simulation results offer critical insights into the comparative performance of the proposed model and the change-point-only model across different values of $\pi_{(s)}$, which denotes the proportion of patients in the stable group. When $\pi_{(s)} = 0$, implying that all subjects are assumed to experience a change point, both the proposed model (i.e., the full model) and the change-point-only model yield similar performance metrics for the parameters associated with the change-point group. Both models exhibit low bias and maintain coverage probabilities close to the nominal 95\% level for the parameters related to the subjects in the change-point group, indicating appropriate model specification and reliable parameter estimation for both models. Since no subjects belong to the stable group under this scenario, the parameters governing the stable group and $\pi_{(s)}$ itself are inherently non-identifiable. Consequently, their estimation results are neither meaningful nor relevant for evaluation in this setting. When the value of $\pi_{(s)}$ increases, the change-point-only model, not accounting for the stable group, displays increasing bias and reduced coverage probabilities. This decline in performance suggests model mis-specification, as it inaccurately attributes all variations in TB to a change point, ignoring those patients whose conditions remain stable. Conversely, the full model, which classifies patients into change-point or stable groups based on the observed data, consistently maintains low bias and meets the 95\% coverage target under different \(\pi_{(s)}\) values. This adaptability underscores the model's effectiveness in capturing the varied dynamics within the data, thereby providing a more accurate and reliable framework for individual outcome prediction.

To further assess the model’s performance in estimating the marginal longitudinal trajectory, we present the corresponding results in the bottom part of Figure \ref{fig:sim_res}. These results clearly show that our proposed full model consistently outperforms the change-point-only model in terms of bias, MSE, and coverage probability. For both settings of $\pi_{(s)} = 0.2$ and $\pi_{(s)} = 0.4$, the full model exhibits lower bias and MSE across all time points, highlighting the advantage of incorporating the complete set of model components for improved estimation accuracy. Notably, our model remains robust even at larger time points, with bias and MSE remaining relatively stable compared to earlier periods, whereas the performance of the change-point-only model deteriorates markedly over time. This suggests that, although both models experience some degradation as time increases---an expected outcome due to the compounding of parameter estimation errors---our full model continues to yield reliable and accurate estimates throughout the entire follow-up period. In addition, the full model consistently achieves coverage probabilities close to the nominal 95\% level, whereas the change-point-only model exhibits significantly lower coverage, particularly at later time points. Overall, our model demonstrates superior accuracy and stability, making it a more reliable choice across different time scales.

In scenarios where the divergence between the pre-slope and post-slope parameters is relatively small, specifically when setting $\bm{\mu}_{r} = (0.3, 0.0, -0.3, 0.3)^T$, parameter estimation may not be as reliable as in the case of $\bm{\mu}_{r} = (0.5, 0.0, -0.5, 0.5)^T$, as shown in the top part of Figure \ref{fig:sim_res}. We posit that this is due to the difficulty in distinguishing between the change point group and the stable group when the pre- and post-slope differences are minimal. Nevertheless, our model consistently outperforms the change-point-only model. While some degree of bias in parameter estimation is observed, the estimation of the marginal longitudinal trajectory remains robust, as shown in the bottom part of Figure \ref{fig:sim_res}, underscoring the practical utility of our approach in tracking longitudinal trajectories for TB. This is particularly relevant in the context of the estimand framework \cite{kahan2024estimands}, as emphasized in regulatory guidance from the U.S. Food and Drug Administration (FDA) \cite{FDA_estimand_guidance}. According to this framework, the primary objective of statistical modeling in clinical trials should be the estimation of clinically meaningful quantities, such as the longitudinal trajectory, rather than achieving perfect accuracy in parameter estimation. Therefore, while parameter estimates should be interpreted with caution in real-world data analysis, the robustness of our model in producing stable trajectory estimates underscores its practical applicability, reinforcing its value for TB progression analysis and potential regulatory decision-making.

\section{Application to the EMPOWER study}
\label{sec:real_data}

We apply our method to the subset of the EMPOWER study mentioned in Section \ref{sec:motivate} and provide an examination of overall TB dynamics through estimated population-level change points, pre-slopes, and post-slopes (defined in Section 6.1, Supplementary Note). And the estimations are provided in Section 6.2 in the Supplementary Note. These parameters reveal significant distinctions: the Cemiplimab treatment arm exhibits a markedly later overall change point at 0.556 years (95\% CI: 0.394, 0.674) compared to 0.301 years (95\% CI: 0.247, 0.359) for the Chemotherapy control arm; critically, the 95\% confidence interval for Cemiplimab's change point (lower bound 0.394 years) does not overlap and is entirely above that for Chemotherapy (upper bound 0.359 years), robustly supporting a delay in TB increase for Cemiplimab. Regarding the initial tumor response, the pre-slope for Cemiplimab is -0.174 (95\% CI: -0.217, -0.142), suggesting a slightly more rapid mean initial TB reduction than Chemotherapy's -0.149 (95\% CI: -0.185, -0.116). Most notably, the post-change point TB dynamics differ substantially; Cemiplimab demonstrates a slower TB increase with a post-slope of 0.091 (95\% CI: 0.041, 0.206), whereas Chemotherapy shows a more rapid increase at 0.31 (95\% CI: 0.23, 0.383), with their respective CIs not overlapping. Collectively, these findings suggest Cemiplimab treatment not only delays the onset of tumor regrowth but also considerably attenuates its subsequent rate, indicative of a more sustained control over TB progression compared to chemotherapy.

\begin{figure}[ht]
    \centering
    \includegraphics[width=1.0\linewidth]{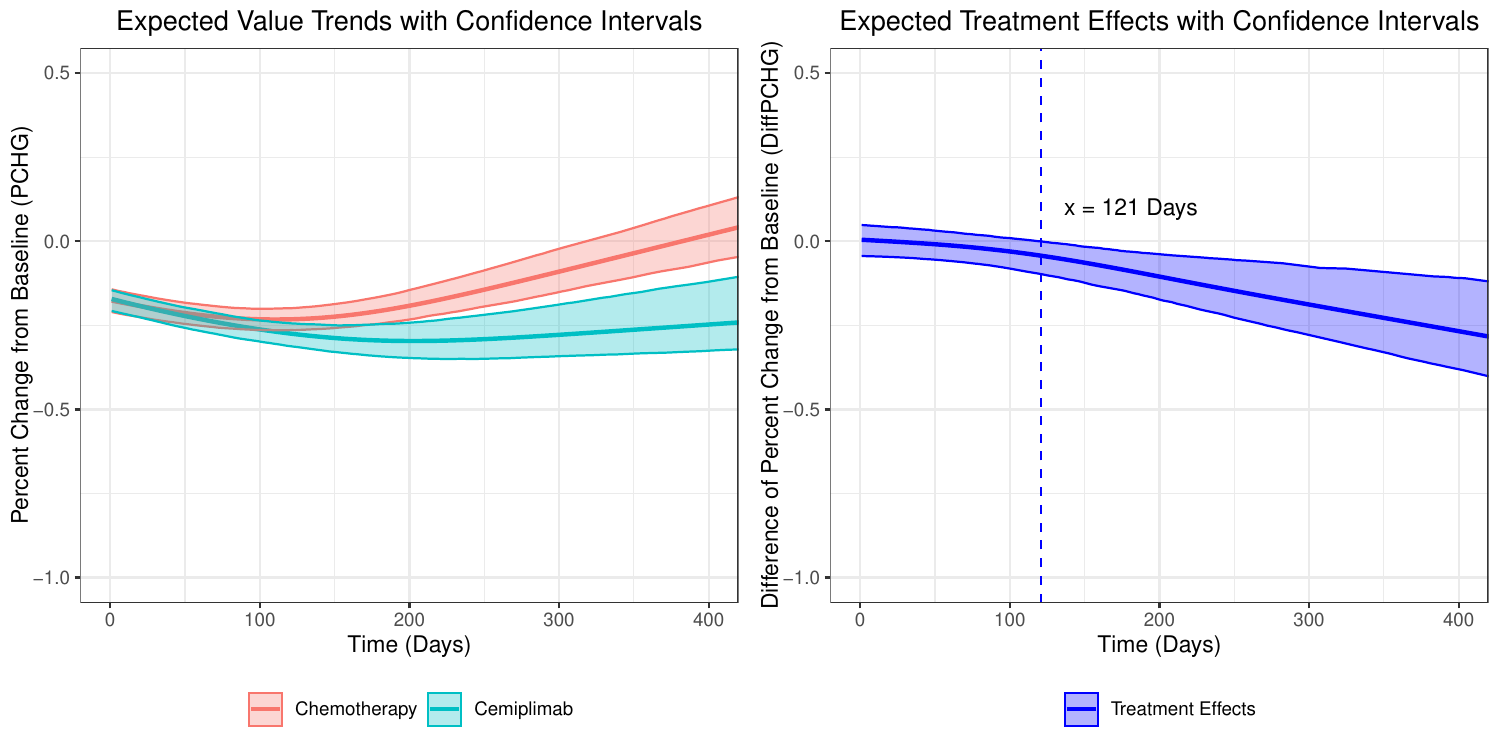}
        \caption{Left: The longitudinal trajectories for both arms, with the vertical dashed line indicating the time point where the CIs of the two arms separate. Right: The estimated treatment effect over time, with the vertical dashed line marking the point where the upper bound of the CI falls below zero, indicating a significant treatment benefit.}
    \label{fig:real}
\end{figure}

In addition to the overall change point and slope parameters, we also examine the expected longitudinal trajectory of TB for both arms. This analysis is particularly valuable as simulations indicate these trajectory estimates are more stable than individual parameter estimates, and such marginal outcomes are of considerable interest within the estimand framework. As depicted in the left panel of Figure~\ref{fig:real}, the expected TB for both arms initially declines from comparable baseline values before subsequently increasing. Crucially, throughout the observation period, the Cemiplimab arm consistently exhibits lower mean TB levels than the Chemotherapy arm, pointing to a sustained treatment effect. The 95\% CIs for the mean TB trajectories of the two arms become distinct after approximately 180 days and remain separated thereafter, despite widening over time, suggesting a meaningful and persistent divergence in TB trajectory. To further quantify this divergence, the treatment effect, defined as the difference in mean TB between the Cemiplimab and Chemotherapy arms, is also estimated (Figure~\ref{fig:real}, right panel). The 95\% CI for this treatment effect falls entirely below zero from approximately 121 days onward, indicating that Cemiplimab achieves statistically significantly lower TB levels compared to Chemotherapy from this time point. This result provides robust, time-dependent evidence of Cemiplimab's efficacy in reducing and controlling TB over the course of the study.

To further investigate treatment effects within specific patient cohorts, subgroup analyses are performed based on PD-L1 expression levels. Patients were categorized into three groups: 50\% $\leq$ PD-L1 $\leq$ 60\% (n=147), 60\% $<$ PD-L1 $<$ 90\% (n=120), and PD-L1 $\geq$ 90\% (n=133). The total of 400 patients across these subgroups corresponds to the subset described in Section \ref{sec:motivate}. The estimated overall change points, pre-slopes, and post-slopes for these subgroups revealed distinct patterns (details in Section~6.2 of the Supplementary Note). Consistent with the full analysis, Cemiplimab demonstrated superiority over Chemotherapy within each PD-L1 subgroup, achieving consistently later mean times to TB increase (overall change point) and slower subsequent tumor regrowth (overall post-slope). Examining the Cemiplimab arm across these PD-L1 expression levels, a clear trend of enhanced efficacy with increasing PD-L1 expression emerges. The overall change point for Cemiplimab progressively extended (from 0.37~years in the 50-60\% PD-L1 group to 0.896~years in the $\geq$90\% group), while the post-change point slope dramatically decreases (from 0.204 to 0.003, respectively, across the same PD-L1 subgroups). These findings indicate increasingly durable tumor control and minimal regrowth in patients with higher PD-L1 expression, aligning with the biological principle of PD-1/PD-L1 targeted therapies \cite{sun2018regulation}, which further demonstrates our method's analytical power to quantify clinically impactful and biologically informative treatment dynamics.

\section{Discussion}
\label{sec:discussion}

This study introduces a novel cure-rate joint model that captures the non-linear dynamics of TB by integrating time-to-event and longitudinal data. The contributions of our model to statistical modeling are threefold. First, our approach explicitly models patients with stable TB and those experiencing a change point, enabling a more flexible representation of disease progression. Second, unlike conventional joint models that often impose linearity on longitudinal trajectories, our framework accommodates individual-specific change points for those in the change-point group, allowing for a more precise characterization of tumor dynamics. Lastly, our model accounts for the dependence of the change point on progression time, leveraging the time-to-event model to enhance inference for the longitudinal component in patients classified within the change-point group. Simulation studies confirm the robustness of our model in estimating the marginal longitudinal trajectory across diverse scenarios, including extreme cases where $\pi_{(s)} = 0$ or when distinguishing between the two groups is particularly challenging. When applied to real-world data from the EMPOWER study, our model effectively differentiates treatment effects on TB progression. The longitudinal trend analysis demonstrates that Cemiplimab consistently maintains lower TB levels over time, providing strong evidence of the model’s ability to quantify differential treatment impacts. Moreover, subgroup analysis reveals that higher PD-L1 expression results to more favorable treatment outcomes, underscoring the model's utility in deriving clinically relevant insights.

The proposed model offers a significant advancement in the analysis of TB dynamics, with several opportunities for further refinement. First, it assumes a log-normal distribution for time-to-event outcomes, which, while computationally efficient, may not always align with real-world data. Second, the current framework is designed for univariate longitudinal outcomes, limiting its ability to capture relationships among multiple biomarkers commonly measured in oncology trials. Additionally, the model does not currently incorporate competing risks, such as distinguishing between progression-related events and treatment-related toxicities, which could enhance its clinical applicability. Moreover, as a frequentist approach, it does not inherently incorporate prior knowledge, an aspect that could be addressed through Bayesian extensions.

Several promising directions could further expand the model’s capabilities. Incorporating more flexible time-to-event distributions, such as Weibull or non-parametric approaches, would allow for greater adaptability to diverse survival data structures. Integrating multiple biomarkers into the longitudinal component could provide a more comprehensive understanding of disease progression. Additionally, accounting for competing risks would improve the model’s applicability in clinical settings where multiple event types must be considered. Exploring Bayesian adaptations could further enhance the framework by enabling the integration of prior knowledge and hierarchical structures, improving inference and uncertainty quantification.

\section*{Acknowledgements}

The authors thank Drs. Frank Seebach, Bo Gao, Siyu Li, and Yuntong Li for their valuable insights regarding the application.

\bibliographystyle{unsrt}
\bibliography{sampbib}
\includepdf[pages=-]{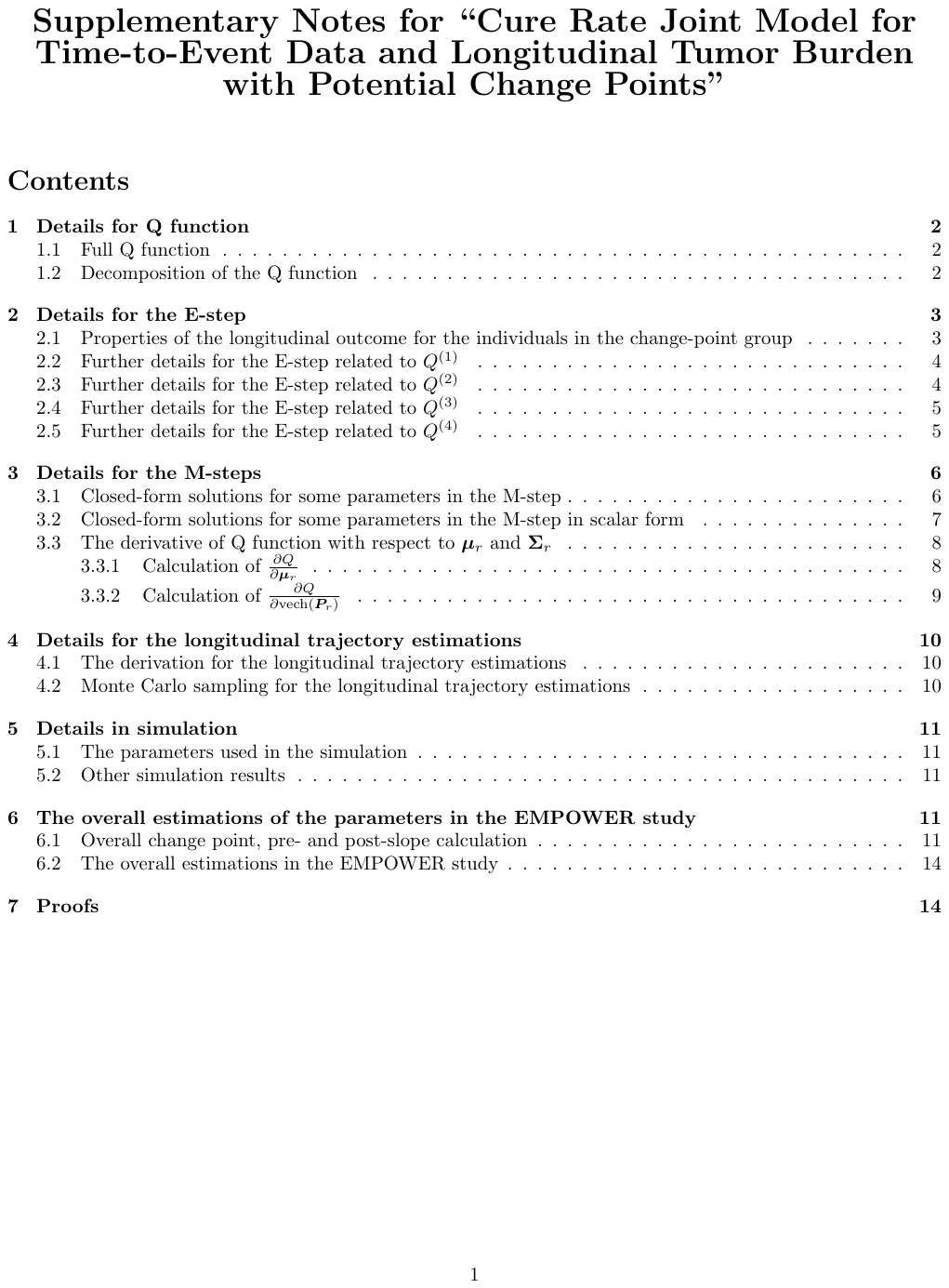} 

\end{document}